\documentclass[epj,final]{svjour}

\usepackage{amsmath}
\usepackage{amssymb}
\usepackage{graphicx}

\DeclareGraphicsExtensions{.eps}
\graphicspath{{./figures/}}
\newlength{\figwidth}

\begin{document}

\title{Interacting electron systems between Fermi leads: \\
effective one-body transmissions and correlation clouds}
\titlerunning{Correlation clouds around interacting electron systems}

\author{%
Rafael A.\ Molina\inst{1,2} 
\and 
Dietmar Weinmann\inst{3} 
\and 
Jean-Louis Pichard\inst{1,4}
}
\authorrunning{R.A.\ Molina \textit{et al.}}

\institute{%
CEA/DSM, Service de Physique de l'Etat Condens{\'e},
Centre d'Etudes de Saclay, 91191 Gif-sur-Yvette, France
\and 
Max-Planck-Institut f\"ur Physik Komplexer Systeme,
N\"othnitzer Str.\ 38, 01187 Dresden, Germany
\and 
Institut de Physique et Chimie des Mat{\'e}riaux de Strasbourg,
UMR 7504 (CNRS-ULP), 23 rue du Loess, BP 43, 67034 Strasbourg Cedex 2,
France
\and 
Laboratoire de Physique Th\'eorique et Mod\'elisation, 
Universit\'e de Cergy-Pontoise, 95031 Cergy-Pontoise Cedex, France
}

\date{Reference: Eur.\ Phys.\ J.\ B \textbf{48}, 243-247 (2005)}

\abstract{%
In order to extend the Landauer formulation of quantum transport to 
correlated fermions, we consider a spinless system in which charge 
carriers interact, connected to two reservoirs by non-interacting 
one-dimensional leads. We show that the mapping of the embedded 
many-body scatterer onto an effective one-body scatterer with 
interaction-dependent parameters requires to include parts of the 
attached leads where the interacting region induces power law 
correlations. Physically, this gives a dependence of the conductance 
of a mesoscopic scatterer upon the nature of the used leads which is
due to electron interactions inside the scatterer. To show 
this, we consider two identical correlated systems connected by a 
non-interacting lead of length $L_\mathrm{C}$. We demonstrate that 
the effective one-body transmission of the ensemble deviates by an 
amount $A/L_\mathrm{C}$ from the behavior obtained assuming an 
effective one-body description for each element and the combination 
law of scatterers in series. $A$ is maximum for the interaction strength 
$U$ around which the Luttinger liquid becomes a Mott insulator in the 
used model, and vanishes when $U \to 0$ and $U \to \infty$. Analogies 
with the Kondo problem are pointed out.
}

\maketitle


\section{Introduction}
 In Landauer's formulation of quantum transport \cite{Landauer}, 
the measure of the conductance $g$ of a coherent system is formulated as 
a scattering problem between incoherent electron reservoirs. In a
two-probe geometry, the system is connected to two reservoirs via leads. 
For large electron densities, the Coulomb interaction is screened and the 
Coulomb to kinetic energy ratio $r_\mathrm{s}$ is small. One has 
essentially a non-interacting system of Fermi energy $E_\mathrm{F}$, 
where the occupation of the one-body levels is given by a Fermi-Dirac 
distribution at a temperature $T$. The system acts as a one-body scatterer 
and its residual conductance $g(T \to 0)$ is given (in units of $2e^2/h$ 
for single channel leads and spin degeneracy) by the probability 
$\vert t(E_\mathrm{F})\vert^2$ of an electron of energy $E_\mathrm{F}$ to 
be elastically transmitted through it. 

\begin{figure}
\centerline{\includegraphics[width=0.6\figwidth]{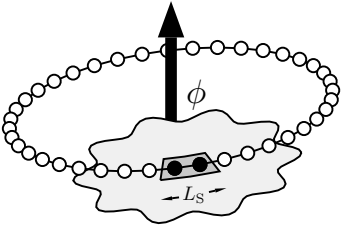}}
\caption{Scheme of the ring pierced by a flux $\phi$ 
used for the embedding method. The correlation cloud induced by 
$L_\mathrm{S}$ interacting sites upon the auxiliary lead is sketched 
in grey.}
\label{FIG1} 
\end{figure}

The problem of describing coherent electronic transport becomes more complex 
in the case where the carrier density is low inside the scatterer, the 
screening ceasing to be effective and the electrons becoming correlated. 
Such situations occur in quantum point contacts of transverse size smaller 
than the Fermi wavelength, where a $0.7\  (2e^2/h)$ structure is observed 
\cite{Thomas}, and can be expected in molecules \cite{Kergueris}, atomic 
chains or contacts \cite{Agrait}, quantum dots where few electrons might 
form a correlated solid, a charge density wave, a Mott insulator, etc. 
In these cases, the electrons are transmitted from one Fermi reservoir 
to another through a many-body scatterer. To extend Landauer's approach  
to such systems, at least for low temperatures and bias voltages, one 
needs to reduce the bare many-body scatterer to an effective one-body 
scatterer with interaction-dependent parameters. This task will be 
hopeless for an isolated system where electrons interact with 
a large interaction strength $U$, but becomes possible when leads where 
electrons do not interact are attached to it. This has been numerically 
demonstrated in previous works \cite{Molina1,Molina2} using the embedding 
method, which allows to extract 
\cite{Molina1,Molina2,Gogolin,Mila,Sushkov,Meden,Rejec} the effective 
coefficient $\vert t(E_\mathrm{F},U)\vert^2$ from the persistent current 
of a large non-interacting ring embedding the many-body scatterer. 
Using the same method, we show that it is not the region where the 
electrons interact which acts as an effective one-body scatterer 
with renormalized parameters, but a larger region where the many-body 
scatterer induces correlations. This problem is somewhat similar 
to the Kondo problem, which can be solved using Wilson's numerical 
renormalization group (NRG) \cite{Hewson}. Instead of using the NRG method, 
we use the density matrix renormalization group (DMRG) method 
\cite{DMRG1,DMRG2} for a non-interacting ring embedding the many-body 
scatterer, as sketched in Fig.\ \ref{FIG1}. 
The modulus of the effective one-body transmission 
amplitude $|t(E_\mathrm{F},U)|$ is obtained from the persistent current 
of the ring extrapolated to infinite lead length, while its phase $\alpha$ 
is given by the Friedel sum rule. 

\begin{figure}
\centerline{\includegraphics[width=0.95\figwidth]{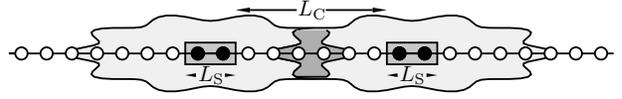}}
\caption{Scheme of the set-up with two identical many-body scatterers 
connected by $L_\mathrm{C}$ sites where the carriers do not interact. 
When $L_\mathrm{C}$ is small, the two correlation clouds sketched in 
grey overlap and the effective one-body scatterer of the ensemble is not 
given by the effective one-body scatterer of each element and the 
combination law of one-body scatterers in series.}
\label{FIG2} 
\end{figure}

After a study of the contained extrapolation, we apply the 
embedding method to determine the effective total transmission coefficient 
$|t_\mathrm{T}(E_\mathrm{F},U)|^2$
of two identical many-body scatterers in series,
connected by a non-interacting lead of size 
$L_\mathrm{C}$ (sketched in Fig.\ \ref{FIG2}). 
The resulting exact value for $|t_\mathrm{T}(E_\mathrm{F},U)|^2$ deviates 
from the one obtained assuming the combination law of one-body 
scatterers in series. This  $U$-dependent deviation is due to 
induced correlations in the attached leads, and its dependence on
$L_\mathrm{C}$ allows to determine the size of the region which 
acts as an effective one-body scatterer. 

\section{Embedding method, extrapolation and correlation cloud}
To study the mapping of a bare many-body scatterer coupled to 
leads onto an effective one-body scatterer with interaction-dependent
coefficients, we take a model of $N$ spinless fermions in a chain of 
$L=L_\mathrm{S} +L_\mathrm{L}$ sites. The Hamiltonian (with even 
$L_\mathrm{L}$) reads
\begin{equation}
\begin{aligned}
H =& -t_\mathrm{h} \sum_{i=2}^{L}(c^\dagger_i c^{\phantom{\dagger}}_{i-1} 
+ c^\dagger_{i-1}c^{\phantom{\dagger}}_i)\\
& + 
U\sum_{i=L_\mathrm{L}/2+2}^{L_\mathrm{L}/2+L_\mathrm{S}}\left[n_i-V_+\right]
\left[n_{i-1}-V_+\right] \, .
\end{aligned}
\label{hamiltonian}
\end{equation}
The hopping amplitude $t_\mathrm{h}=1$ between nearest neighbor sites sets 
the energy scale, $c^{\phantom{\dagger}}_i$ ($c^\dagger_i$) is 
the annihilation (creation) operator at site $i$, and 
$n_i = c^\dagger_i c^{\phantom{\dagger}}_i$. 
The nearest neighbor repulsion $U$ acts upon $L_\mathrm{S}$ 
consecutive sites and gives rise to many-body scattering. 
We take a half-filled model ($N=L/2$), with a potential $V_+=1/2$ being 
due to a positive background charge which exactly compensates the 
repulsion $U$ inside the scatterer. Therefore, our model exhibits 
particle-hole symmetry and a uniform density, without Friedel 
oscillations around the scattering region where the fermions interact. 

The scattering geometry corresponds to two leads of $L_\mathrm{L}/2 \to 
\infty$ sites connected by an interacting scatterer of $L_\mathrm{S}$ sites. 
The electrons do not interact in the leads, a necessary condition for 
having appropriate asymptotic scattering channels in one dimension. 
For the embedding method, we consider the ring geometry sketched in 
Fig.\ \ref{FIG1}, the scatterer being closed on itself via a 
non-interacting lead of $L_\mathrm{L}$ sites. This is achieved 
by adding a hopping term 
$$
-t_\mathrm{h}c^\dagger_1c^{\phantom{\dagger}}_L\exp(i\phi)+\mathrm{h.c.} 
$$
to the Hamiltonian (\ref{hamiltonian}), the flux $\phi$ driving a persistent 
current $J(U)$ in the ring. As the flux dependence of 
$J(U)$ extrapolated to the limit $L_\mathrm{L}\to\infty$ 
demonstrates \cite{Molina2}, the many-body scatterer behaves as an 
effective one-body scatterer, but with an interaction-dependent 
elastic transmission coefficient $\vert t(E_\mathrm{F},U)\vert^2$. 
Instead of using $J(U)$, it is simpler \cite{Molina2} to get 
$\vert t(E_\mathrm{F},U)\vert^2$ from the charge stiffness 
\begin{equation}
D(U,L_\mathrm{S},L)=(-1)^N \frac{L}{2}\big( E_0(U,L_\mathrm{S},L)-
E_{\pi}(U,L_\mathrm{S},L)\big) \  ,
\label{eq:stiffness}
\end{equation}
where $E_0(U,L_\mathrm{S},L)-E_{\pi}(U,L_\mathrm{S},L)$ is the change 
of the ground-state energy from periodic to antiperiodic boundary 
conditions. $D(U,L_\mathrm{S},L)$ is obtained by the DMRG implementation  
for real Hamiltonians, which can be used to study with a great accuracy 
systems as large as $L=120$ sites with $N=60$ particles. In the limit 
$L_\mathrm{L}\to\infty$, one gets the modulus 
\begin{equation}
\left|t(E_\mathrm{F},U)\right|=\sin \left( \frac{\pi}{2}
\frac{D_{\infty}(U,L_\mathrm{S})}{D_{\infty}(U=0,L_\mathrm{S})}\right)
\label{eq:stiff}
\end{equation}
of the transmission amplitude through the scatterer of $L_\mathrm{S}$ sites,
$D_{\infty}(U=0,L_\mathrm{S})$ being the charge stiffness of the same 
ring for $U=0$. 


To take the limit $L_\mathrm{L}\to\infty$ is one of the key points of 
the embedding method. This extrapolation is also required for pure 
one-body scattering, where the finite size corrections to 
formula (\ref{eq:stiff}) can be expanded \cite{Molina2} in powers of 
$1/L$. For many-body scattering, the DMRG study gives an empirical 
scaling law \cite{Molina1,Molina2} 
\begin{equation}
D(U,L_\mathrm{S},L) = D_{\infty}(U,L_\mathrm{S})
\exp{\left(\frac{C(U,L_\mathrm{S})}{L}\right)} 
\label{eq:conv}
\end{equation} 
obtained for large $L_\mathrm{L}$ and small $L_\mathrm{S}$, 
which allows to determine the asymptotic value $D_{\infty}(U,L_\mathrm{S})$ 
necessary to obtain $\left|t(E_\mathrm{F},U)\right|^2$. Expanding this 
scaling law gives 
\begin{equation}
D(U,L_\mathrm{S},L)-D_{\infty}(U,L_\mathrm{S}) \approx 
\frac{B(U,L_\mathrm{S})}{L}
\end{equation}
when $L$ is large enough, where 
\begin{equation}
B(U,L_\mathrm{S})=C(U,L_\mathrm{S})D_{\infty}(U,L_\mathrm{S})\, .
\end{equation} 
This is a power law decay, and not an exponential decay with a 
characteristic scale above which the finite size correction can be 
neglected. Numerical data show that $B(U,L_\mathrm{S})$ is 
important for intermediate interaction strengths $U$.
But in the limits $U\to 0$ (no scattering, total transmission) and 
$U \to \infty$ (total reflection, 
$D_{\infty}(U \to \infty,L_\mathrm{S}) \to 0$),
the finite size corrections vanish and $B(U,L_\mathrm{S}) \to 0$. 

However, since adding one-body potentials in the region of the $L_\mathrm{S}$ 
sites yields a finite size correction to $D(U,L_\mathrm{S},L)$ 
even when $U \to 0$, the interpretation of these corrections is not 
straightforward. They do not depend only on the correlations 
induced in the attached lead by the interaction acting inside the 
scatterer, but also on more trivial one-body aspects. 

\section{Combination of two many-body scatterers in series}
A more direct approach, where the finite size effects 
are only due to many-body correlations, consists in taking two 
identical scatterers connected by a scattering-free lead of 
size $L_\mathrm{C}$ in which the electrons do not interact, as sketched 
in Fig.\ \ref{FIG2}. Since the scattering channels begin at the first 
attached sites of the leads when $U=0$, there is a simple combination 
law for one-body scatterers in series. To study how this combination 
law is broken with increasing $U$ when $L_\mathrm{C}$ is small allows 
to show that the size of the effective elastic scatterer is larger than 
the region where the carriers interact. When $U \neq 0$, the scattering 
channels begin only asymptotically far from the many-body scatterer. 


\begin{figure}
\centerline{\includegraphics[height=0.6\figwidth]{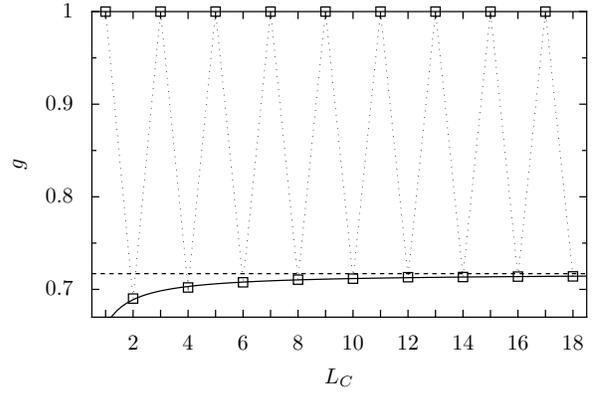}}
\caption{Conductance $g(L_\mathrm{C})$ of the set-up sketched in 
Fig.\ \ref{FIG2} with $L_\mathrm{S}=2$, for $U=1$. 
The points are obtained with the embedding method for the 
ensemble. The dashed line at $g=0.717$ 
gives the approximate value obtained from Eq.\ (\ref{eq:combeven}), 
with $|t|$ obtained by the embedding method for a single scatterer. 
The solid line is a fit with the form 
$g(L_\mathrm{C})=0.7174-0.057/L_\mathrm{C}$.
}
\label{FIG3} 
\end{figure}
\begin{figure}
\centerline{\includegraphics[height=0.6\figwidth]{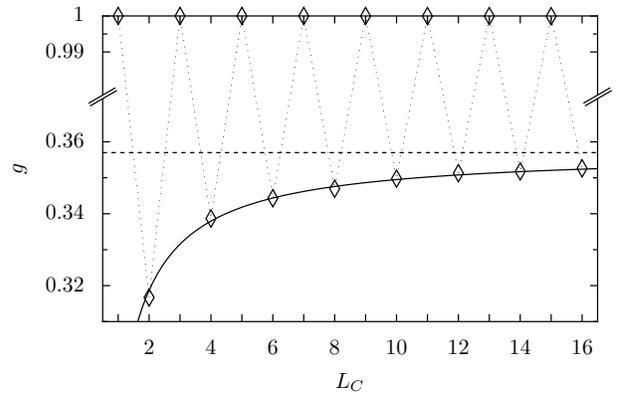}}
\caption{Conductance $g(L_\mathrm{C})$ as in Fig.\ \ref{FIG3}, but for 
$U=2$. The dashed line at $g=0.357$ represents the value yielded  
by Eq.\ (\ref{eq:combeven}), and the solid line is the fit 
$g(L_\mathrm{C})=0.3572-0.077/L_\mathrm{C}$.
}
\label{FIG4} 
\end{figure}
Without interaction, a scatterer can be described at energy $E_\mathrm{F}$ 
by a unitary scattering matrix $S_\mathrm{S}$, written in 
terms of its reflection and transmission amplitudes $r,r'$ and $t,t'$ as
\begin{equation}
S_\mathrm{S}=\left(\begin{array}{cc} 
r & t' \cr t & r' \end{array}
\right)\, .
\end{equation}
The scatterer being symmetric upon time reversal, one 
has $t=t'$, while $r=r'$ if the scatterer is symmetric upon space 
inversion. The transfer matrix  $M_\mathrm{S}$ (giving the flux 
amplitudes at the right side in terms of the flux amplitudes at 
the left side) reads 
\begin{equation}
M_\mathrm{S}=\left(\begin{array}{cc} 1/t & r/t \cr r^*/t^* & 1/t^* \end{array}
\right)\, .
\end{equation}
The total Hamiltonian and the parity operator can be simultaneously 
diagonalized if one has inversion symmetry, to give even and odd 
standing-wave solutions which can be written as 
$\psi_i^0=\cos(ki+\delta_0)$ and
$\psi_i^1=\sin(ki+\delta_1)$
at the right side of the scatterer, and 
$\psi_i^0=\cos(ki-\delta_0)$ and
$\psi_i^1=\sin(ki-\delta_1)$ 
at its left side. The two phase shifts 
$\delta_0$ and $\delta_1$ are related \cite{Lipkin} to $t$ and $r$ by 
\begin{equation}
\begin{aligned}
t&=(\exp (2i\delta_0)+\exp (2i\delta_1))/2\, , \\
r&=(\exp (2i\delta_0)-\exp (2i\delta_1))/2\, . 
\end{aligned}
\end{equation}
Due to symmetries, $S_\mathrm{S}$ or 
$M_\mathrm{S}$ have only two free parameters: the modulus 
$|t|=\cos(\delta_0-\delta_1)$ and the phase $\alpha=\delta_0+\delta_1$ 
of the transmission amplitude $t$, the unitarity of $S_\mathrm{S}$ 
($|t|^2+|r|^2=1$ and $r/r^*=-t/t^*$) giving $r$. We can determine $|t|$ 
by the embedding method. The Friedel sum rule \cite{Friedel} gives $\alpha$. 
If one introduces a scatterer with inversion symmetry in the central 
region of a scattering free lead, this rule states \cite{Langer} that 
\begin{equation}
\alpha=\delta_0+\delta_1=\pi N_\mathrm{f}
\end{equation}
for spinless fermions in one dimension. $N_\mathrm{f}$ is the number 
of displaced fermions when the scatterer is introduced in the central 
region. For a uniform filling factor $\nu=1/2$, 
$N_\mathrm{f}=L_\mathrm{S}/2$, and the phase $\alpha$ reads   
\begin{equation}
\alpha=\pi N_\mathrm{f}= \frac{\pi L_\mathrm{S}}{2}= k_\mathrm{F}
L_\mathrm{S}\, ,
\label{eq:phase}
\end{equation}
where $k_\mathrm{F}=\pi/2$ is the Fermi wave number. For the spinless case in 
one dimension with a uniform density, this simply means that the transmitted 
wave has $N_\mathrm{f}$ changes of sign when one transfers a fermion through 
a scatterer containing $N_\mathrm{f}$ others. This is obvious for $U=0$ as 
well as for $U\neq 0$. Using the same rule, the ideal ballistic lead of 
$L_\mathrm{C}$ sites has a modulus $|t(L_\mathrm{C})|=1$ and a phase 
$\alpha(L_\mathrm{C})=k_\mathrm{F} L_\mathrm{C}$. Its transfer matrix reads
\begin{equation}
M_\mathrm{C}=\left(\begin{array}{cc} 
e^{-ik_\mathrm{F}L_\mathrm{C}} & 0 \cr 0 & e^{ik_\mathrm{F}L_\mathrm{C}} 
\end{array}
\right)\, .
\end{equation}
The combination law of one-body scatterers in series being  
a simple matrix multiplication for the transfer matrices, 
the total transfer matrix $M_\mathrm{T}(E_\mathrm{F})$ of the 
ensemble is given by 
\begin{equation}
M_\mathrm{T}=M_\mathrm{S}\cdot M_\mathrm{C} 
\cdot M_\mathrm{S}\, . 
\end{equation}
For the total transmission coefficient 
$|t_\mathrm{T}|^2$ through the ensemble, expressed in terms of the 
transmission $t$ of each element and of $L_\mathrm{C}$, this gives 
\begin{equation}
|t_\mathrm{T}|^2=\frac{|t|^4}{2(1-|t|^2)
(1+\cos{(2k_\mathrm{F}L_\mathrm{C}-2\alpha)})+|t|^4}\, .
\label{eq:combination}
\end{equation}
Since $|t|=1$ when $L_\mathrm{S}$ is odd \cite{Molina1,Molina3}, we 
consider only even values of $L_\mathrm{S}$. Taking 
$\alpha=\pi L_\mathrm{S}/2$ gives the Landauer conductance 
$g=|t_\mathrm{T}|^2=1$ if $L_\mathrm{C}$ is odd and 
\begin{equation}\label{eq:combeven}
g=\frac{|t|^4}{\left(|t|^2-2\right)^2}
\end{equation} 
for $L_\mathrm{C}$ even.

\begin{figure}
\centerline{\includegraphics[width=0.85\figwidth]{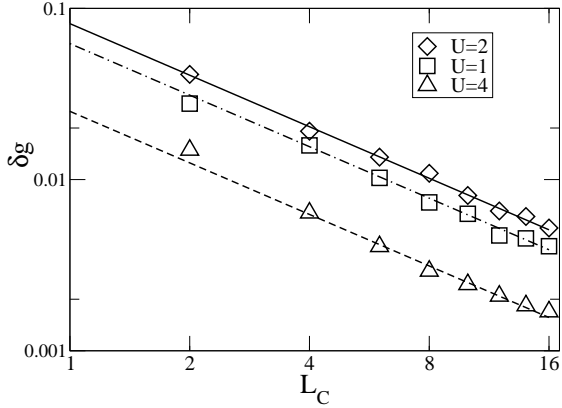}}
\caption{Error $\delta g$ made when using the combination law 
of Eq.\ (\ref{eq:combination}) for having the conductance 
of the set-up sketched in Fig.\ \ref{FIG2} with $L_\mathrm{S}=2$ and 
different values of $U$, as a function of an even number $L_\mathrm{C}$ 
of connecting sites. The lines give an $A(U,L_\mathrm{S})/L_\mathrm{C}$ 
fit.}   
\label{FIG5} 
\end{figure}
\section{Correlation-induced deviations from the non-interacting combination law}
Fig.\ \ref{FIG3} and \ref{FIG4} show the conductance $g$ for two 
scatterers of $L_\mathrm{S}=2$ sites in series, as a function of the length 
$L_\mathrm{C}$ of the coupling lead. 
The data points are 
directly obtained from the embedding method, without assuming a combination 
law for scatterers in series. Resonances with $g=1$ occur for odd
$L_\mathrm{C}=\{1,3,5,\dots \}$. For even $L_\mathrm{C}$, the
dashed lines represent the $L_\mathrm{C}$-independent values 
$|t|^4/(|t|^2-2)^{2}$ implied by Eq.\ (\ref{eq:combination}), 
the coefficient $|t|$ being obtained using the embedding method for a 
single scatterer. Within the accuracy of the extrapolation procedures 
required for having the transmission $|t|$ of an individual scatterer 
and the total conductance $g$, the result of (\ref{eq:combination}) 
gives the correct value when $L_\mathrm{C}\to \infty$, but overestimates 
$g$ for small even values of $L_\mathrm{C}$. The difference 
\begin{equation}
\delta g (L_\mathrm{C}) = g(L_\mathrm{C}\to \infty)-g(L_\mathrm{C})
\end{equation}
is shown in Fig.\ \ref{FIG5} for even $L_\mathrm{C}$ at different 
values of $U$. 
\begin{figure}
\centerline{\includegraphics[width=0.85\figwidth]{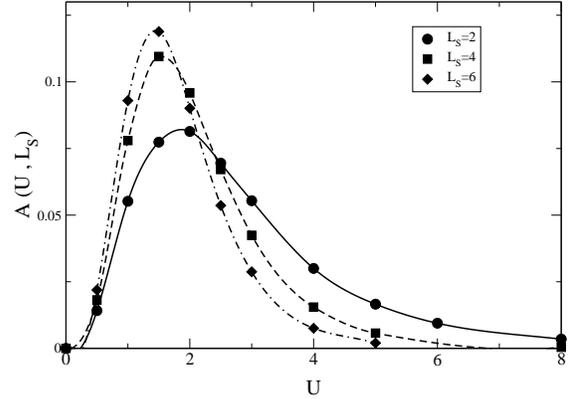}}
\caption{Amplitude $A(U,L_\mathrm{S})$ of the fits shown in Fig.\
  \ref{FIG5} as a function of $U$ for 
different values of $L_\mathrm{S}$.}   
\label{FIG6} 
\end{figure}
For even $L_\mathrm{C}$, $\delta g(L_\mathrm{C})$ decays as a function 
of $L_\mathrm{C}$ as 
\begin{equation}
\delta g(L_\mathrm{C}) \approx
\frac{A(U,L_\mathrm{S})}{L_\mathrm{C}}\, ,
\end{equation}
with an amplitude $A(U,L_\mathrm{S})$ which is shown in Fig.\
\ref{FIG6} as a function of the interaction strength $U$. 
This $1/L_\mathrm{C}$ decay is reminiscent of the $1/L$ decay 
characterizing $D(U,L_\mathrm{S},L)-D_{\infty}(U,L_\mathrm{S})$ 
and of the screening at large distances (larger than the Thomas-Fermi 
screening length) of the potential of a point charge by non-interacting 
electrons (Friedel oscillations, RKKY interactions \dots) in one dimension. 
This suggests that the decay could be faster for leads of higher 
dimensions ($1/L_\mathrm{C}^d$ decay in $d$ dimensions). The amplitude 
$A(U,L_\mathrm{S}) \rightarrow 0$ when $U \rightarrow 0$ (one-body
scatterers) and when $U \rightarrow \infty$. In this latter limit, 
the scatterers become decoupled from the leads, the energy for an electron 
to enter or to leave a scatterer being $\propto U$. $A(U)$ is maximum near 
$U=2$, a value where in the thermodynamic limit 
$L_\mathrm{S}\rightarrow\infty$ the Luttinger liquid becomes 
\cite{Giamarchi} a Mott insulator for spinless fermions. 

For all odd values of $L_\mathrm{C}$, the data for the total
transmission coincide with the value $g=1$ obtained from 
Eq.\ (\ref{eq:combination}), and $\delta g(L_\mathrm{C})=0$.
The even-odd dependence on the parity of $L_\mathrm{C}$ shows 
that the convergences of the phase $\alpha$ and the modulus $|t|$ of 
the effective scatterer are characterized by different scales. One has
$\alpha=\pi N_\mathrm{f}$ across the scatterer, directly on a scale 
$L_\mathrm{S}$, independently of $U$, while $|t|$ reaches its asymptotic 
value on a much larger scale. This is not surprising since $\alpha$ depends 
on the mean density, while $|t|$ depends on the  correlations of its 
fluctuations. In our model with a compensating background charge, the mean 
density does not exhibit Friedel oscillations. Let us underline that the 
correlation clouds which have to be included with the many-body scatterer 
to form the effective one-body scatterer must not be confused with the 
screening clouds characterizing the charge density. 


\section{Discussion of the relation to the Kondo problem} 
To obtain the effective one-body matrix $S(E_\mathrm{F})$ of a 
correlated system of spinless fermions is a problem which displays 
a certain similarity with the Kondo problem of a spin degree of freedom 
surrounded by a metallic host. In the two cases, it is crucial to 
couple the many-body system to non-interacting conduction electrons. 
For the Kondo problem, the original $3d$ model can be mapped onto a 
$1d$ lattice without interaction embedding a Hubbard impurity. 
In Wilson's renormalization group transformations 
\cite{Hewson}, the embedded many-body Hamiltonian is progressively 
mapped onto an effective one-body Hamiltonian describing the low energy 
states. In this transformation, the coupling between different length 
scales is taken into account progressively, working out from the impurity 
to the longer length scales and lower energies. The states at sites near 
the impurity involve conduction states spanning the full band width 
$2t_\mathrm{h}$, while the states located far from the impurity
involve conduction states near the Fermi level, with a progressively 
reduced band width. This NRG method has been used recently \cite{Oguri} to 
calculate the effective one-body Hamiltonian of a few Hubbard sites embedded 
in a non-interacting chain, and the corresponding phase shifts. 

In our spinless case, the embedded system would give rise to inelastic 
and elastic scattering if it was in the vacuum. Due to 
the attached non-interacting leads inelastic processes become 
progressively blocked by a Fermi vacuum which eventually takes place 
in the leads, at a large distance from the scatterer. 
Our DMRG study leads us to a similar conclusion as NRG studies of impurities 
with spin. It shows that the effective one-body elastic scatterer necessary 
for extending the Landauer formulation of coherent transport to correlated 
fermion systems must include parts of the attached leads where the 
interacting region induces power law correlations. Physically, this 
gives a dependence of the conductance of a mesoscopic scatterer upon 
the nature of the used leads which depends on the strength of the 
interactions inside the scatterer. Eventually, let us mention that the 
vertex corrections due to inelastic scattering vanish \cite{oguri97} when 
$T \to 0$, in a perturbative approach to the Kubo conductance of an 
interacting region embedded between semi-infinite leads. While this 
agrees with our findings for $L_\mathrm{L} \to \infty$, it is likely that 
these corrections do not vanish when $L_\mathrm{L}$ is finite, and exhibit 
similar power-law decays as $L_\mathrm{L}$ increases. 

\section{Acknowledgments}
We thank  Y.\ Asada, G.-L.\ Ingold, R.A.\ Jalabert, O.\ Sushkov and G.\ 
Vasseur for stimulating discussions, and P.\ Schmitteckert for his DMRG code. 
R.A.\ Molina acknowledges the financial support provided through 
the European Community's Human Potential Program under contract 
HPRN-CT-2000-00144.

\end{document}